\documentclass[twocolumn,tighten,times]{aastex631}
\usepackage{amsmath}
\usepackage{bm}

\def\ihep{Key Laboratory for Particle Astrophysics, Institute of High Energy Physics, Chinese Academy of Sciences, 19B Yuquan Road, Beijing 100049, China; \href{mailto:liyanrong@ihep.ac.cn}{liyanrong@mail.ihep.ac.cn}, \href{mailto:wangjm@mail.ihep.ac.cn}{wangjm@mail.ihep.ac.cn}}

\def\naoc{National Astronomical Observatory of China, 20A Datun Road, Beijing 100020, China}

\def\UCASastro{School of Astronomy and Space Sciences, University of Chinese Academy of Sciences, Beijing 100049, China}

\begin{document}

\title{\bf \large Radial-dependent Responsivity of Broad-line Regions in Active Galactic Nuclei: Observational Consequences
for Reverberation Mapping and Black Hole Mass Measurements}
\shorttitle{Radial-dependent Echoes of BLRs in AGNs}
\shortauthors{Li \& Wang}

\author[0000-0001-5841-9179]{Yan-Rong Li}
\affiliation{\ihep}
\author[0000-0001-9449-9268]{Jian-Min Wang}
\affiliation{\ihep}
\affiliation{\UCASastro}
\affiliation{\naoc}

\begin{abstract}
The reverberation mapping (RM) technique has seen wide applications in probing geometry and kinematics of broad-line regions (BLRs) and measuring masses of supermassive black holes (SMBHs) in active galactic nuclei. However, the key quantities in RM analysis like emissivity, responsivity, transfer functions, and mean and root-mean-square (RMS) spectra are fragmentally defined in the literature and largely lack a unified formulation. Here, we establish a rigorous framework for BLR RM and include a locally dependent responsivity according to photoionization calculations. The mean and RMS spectra are analytically expressed with emissivity- and responsivity-weighted transfer functions, respectively. We demonstrate that the RMS spectrum is proportional to the responsivity-weighted transfer function only when the continuum variation timescale is much longer than the typical extension in time delay of the BLR, otherwise, biases arise in the obtained RMS line widths. The long-standing phenomenon as to the different shapes between mean and RMS spectra can be explained by a radial-increasing responsivity of BLRs.
The debate on the choice of emission line widths for SMBH mass measurements is explored and the virial factors are suggested to also depend on the luminosity states, in addition to the geometry and kinematics of BLRs.
\end{abstract}

\section{Introduction}
The presence of broad emission lines with widths of several thousand kilometers per second is one characteristic feature of active galactic nuclei (AGNs). Those emission lines stem from the so-called broad-line region (BLR) surrounding the central supermassive black hole (SMBH). It was long suspected that the deep gravitational potential of the SMBH is responsible for such broad line widths, an imprint of fast, probably virialized motion of the ionized BLR gas (e.g., \citealt{Woltjer1959}). However, short variation timescales of broad emission lines point to a compact BLR size, at a level of light-days to light-months, meaning that the exact BLR size is not trivial to measure considering the limited spatial resolution of existing facilities. The advent of reverberation mapping (RM) technique ushered in a new pathway to directly measure BLR sizes as well as diagnose geometry and kinematics of BLRs (\citealt{Bahcall1972, Blandford1982}). The ensuing decades of RM efforts finally led to identification of the evidence for the expected virial relationship (\citealt{{Peterson1999, Peterson2000}}), paving the way for RM-based SMBH mass measurements (e.g., see the review of \citealt{Peterson2014}). The establishment of the relationship between BLR sizes and AGN optical luminosities from RM campaigns (\citealt{Kaspi2000, Bentz2013}) further spawned the single-epoch SMBH mass estimate, reinforcing the central importance of RM for large-scale SMBH demography (e.g., \citealt{Li2011, Shen2012, Schulze2015, Rakshit2020}) and cosmological evolution across the universe (e.g., \citealt{Wang2009, Li2012, Shankar2013, Tucci2017}).

Albeit with these remarkable achievements, there remain several important issues not well understood in RM studies. Firstly, it is long known that the observed mean and root-mean-square (RMS) spectra of broad emission lines in an RM campaign are not always the same in shape, which reflects that the whole BLR does not respond to the continuum variations uniformly. Instead, some parts must be more responsive than the others. If using the terminology ``responsivity'', which quantifies the relative change of line emissivity with respect to that of the incident continuum (\citealt{Goad1993}), the responsivity of the BLR is not a global constant, but locally dependent. This had been well known from photoionization calculations (\citealt{Goad1993, Korista2004, Goad2014, Goad2015}), however, it remains unresolved how the mean and RMS spectra quantitatively depend on the key quantities in RM analysis, such as emissivity, responsivity, and emissivity- and responsivity-weighted transfer functions. Moreover, these key quantifies were fragmentally defined in previous studies and largely lacks a unified formulation.

Secondly, RM-based SMBH mass measurements rely on a scaling factor, so called virial factor, to convert the virial product, computed from a combination of the emission line width and BLR size, to the true SMBH mass (e.g., \citealt{Peterson2014}). The commonly used line width measures  in the literature include the full width at half maximum (FWHM) and line dispersion from either mean or RMS spectra. There is a broad, still ongoing debate as to the optimal choice of line width for accurately estimating SMBH masses (e.g. \citealt{Dalla2020}). A practically feasible framework would shed light on this debate and allow one to explore observational factors affecting virial factors and SMBH mass measurements from a theoretical perspective.

Thirdly, the BLR dynamical modeling approach provides a direct way to measure SMBH masses and thereby determine virial factors of individual AGNs (\citealt{Pancoast2011, Pancoast2014, Li2013, Li2018}). This complements the traditional approach in which statistically averaged virial factors are calibrated by the correlations between SMBH masses and host galaxy properties well established in quiescent galaxies (e.g., \citealt{Onken2004, Ho2014, Yan2024}). All previous dynamical modeling works presumed a globally linear response of the BLR so that the whole BLR is forced to respond uniformly.  It is clear that such modeling cannot appropriately recover the differences in shape between mean and RMS spectra. \cite{Li2013} and \cite{Li2018} included a non-linearity parameter for BLR responses but still set it spatially constant, thereby subject to the same issue.

Fourthly, spectroastrometry (SA) with GRAVITY instrument on the Very Large Telescope Interferometer (VLTI; \citealt{GRAVITY2017}) enables directly resolving BLRs of bright AGNs at an angular resolution down to $\sim$10 micro-arcseconds  (\citealt{GRAVITY2018, GRAVITY2024, Abuter2024}).  A joint analysis of SA and RM (hereafter abbreviated to SARM) observation data (\citealt{Wang2020, Li2022, Li2023}) can not only more thoroughly probe BLR properties, but also constitutes a geometric probe for cosmic distances based on the simple notion that the angular size conveyed from SA and physical size of the BLR from RM can determine the angular-size distance (\citealt{Elvis2002, Rakshit2015, Wang2020}).
However, a raised subtle issue is that SA measures the BLR's photocenter offset as a function of wavelength, related to the emissivity-weighted size, while RM measures the time delay between variations of the continuum and emission line, which is indeed related to the responsivity-weighted BLR size. These two sizes are not always identical, evident from photoionization calculations (\citealt{Goad1993, Zhang2021}) as well as the observed shape differences between mean and RMS spectra mentioned above. A formulation self-consistently taking into account emissivity and responsivity weighting of BLR responses is therefore crucial for diagnosing BLR kinematics and reinforcing the cosmic potential of SARM analysis.

Motivated by all of the above, this paper is devoted to establish a rigorous framework to delineate BLR reverberation with locally dependent responsivity and explore its observational implications. The paper is organized as follows. Section~\ref{sec_form} presents a formulation for RM and derives expressions for mean and RMS spectra. Section~\ref{sec_example} constructs a simple BLR dynamical model and performs simulations to showcase the generated transfer functions and line widths of mean and RMS spectra. Section~\ref{sec_implication} illustrates how a radial-dependent responsivity affects the line widths and BLR sizes, and their implications for SMBH mass measurements and BLR dynamical modeling. The conclusions are summarized in Section~\ref{sec_conclusion}.

\begin{deluxetable*}{llc}
\tablecolumns{3}
\tabletypesize{\footnotesize}
\tabcaption{\centering The Description of Major Notations. \label{tab_variable}}
\tablehead{
\colhead{Notation~~~~~~~~~~~~~~~~} & \colhead{Description} & \colhead{~~~~~~~~~~~~~~~~Equations~~~~~~~~~~~~~~~~}
}
\startdata
$\eta$               & Responsivity                                       & \ref{eqn_resp0}, \ref{eqn_resp2}\\
$F_c(t)$             & Continuum flux density at time $t$                 & \ref{eqn_fc}\\
$\bar F_c$           & Mean continuum flux density                        & \ref{eqn_fc}\\
$\Delta F_c(t)$      & Variable component of $F_c(t)$                     & \ref{eqn_fc}\\
$F_l(t, v)$          & Emission line profile at time $t$ and velocity $v$ & \ref{eqn_fl}, \ref{eqn_rm_org}, \ref{eqn_rm_full}\\
$\bar F_l(v)$        & Mean component of $F_l(t, v)$                      & \ref{eqn_rm_mean}, \ref{eqn_mean} \\
$\Delta F_l(v, t)$   & Variable component of  $F_l(t, v)$                 & \ref{eqn_rm_var}, \ref{eqn_var}  \\
$\Psi_{\rm r}(\tau, v)$    & Responsivity-weighted transfer function            & \ref{eqn_tf_resp}\\
$\Psi_{\rm r}(v)$          & Delay integral of $\Psi_{\rm r}(\tau, v)$                & \ref{eqn_tf_resp_v}\\
$\widehat{\Psi}_{\rm r}(v)$& Delay integral of $\Psi_{\rm r}^2(\tau, v)$              & \ref{eqn_rms2}\\
$\Psi_{\rm e}(\tau, v)$    & Emissivity-weighted transfer function              & \ref{eqn_tf_emis}\\
$\Psi_{\rm e}(v)$          & Delay integral of $\Psi_{\rm e}(\tau, v)$                & \ref{eqn_tf_emis_v}\\
$S_c(\Delta t)$      & Continuum covariance function                      & \ref{eqn_sc}
\enddata
\end{deluxetable*}

\section{Formulation}\label{sec_form}
We start with introducing basic terminology in Section~\ref{sec_term} and then present a generic formulation for BLR RM using the linearized response model in Section~\ref{sec_rm}.
Based on this formulation, we derive rigorous expressions for mean and RMS spectra using transfer functions of the BLR in Section~\ref{sec_rms}.
The detailed derivations for the expressions of BLR reverberation and RMS spectrum are presented in Appendix~\ref{sec_app_rm} and \ref{sec_app_rms}, respectively. The major notations and their descriptions are listed in Table~\ref{tab_variable}.

Throughout the paper, we do not distinguish between the continuum ionizing the BLR and that realistically observed in RM and simply assume that their variations are closely correlated. This is supported by multi-band continuum RM of AGNs, which reveals strong inter-band correlations from UV to optical (e.g., \citealt{Edelson2019}).

\subsection{Basic Terminology}\label{sec_term}
In an RM campaign, the observables include a series of continuum flux $F_{c}(t)$ and emission line profiles $F_l(t, v)$ over a time period $T$.
In general, the continuum flux can be decomposed into the constant and variable components,
\begin{equation}\label{eqn_fc}
F_c(t) = \bar F_c + \Delta F_c(t).
\end{equation}
Similarly, the emission line profile can also be written
\begin{equation}
F_l(t, v) = \bar F_l(v) + \Delta F_l(t, v).
\label{eqn_fl}
\end{equation}
The {\it mean spectrum} is then computed as
\begin{equation}
{\rm Mean} = \bar F_l(v) = \frac{1}{T} \int_T F_l(t, v) d t.
\end{equation}
and the {\it RMS spectrum} is computed as
\begin{eqnarray}\label{eqn_rms_def}
{\rm RMS} = \left[\frac{1}{T} \int_T \left[\Delta F_l(t, v)\right]^2 d t\right]^{1/2}.
\end{eqnarray}
Here, the integrals can also be alternatively rewritten in a form of discrete summation over all epochs under investigation.
In practice, there are other variant forms of definitions for mean and RMS spectra by including weights to each spectrum to account for different data quality (e.g., \citealt{Park2012}). Those definitions do not change the basic results of this work.

From the constant and variable components of the continuum and emission line fluxes, one can introduce two transfer functions as
\begin{equation}
\bar F_l(v) = \bar F_c \int \Psi_{\rm e}(v, \tau)d\tau =  \bar F_c \Psi_{\rm e}(v),
\label{eqn_rm_mean}
\end{equation}
and
\begin{equation}
\Delta F_l(v, t) = \int \Psi_{\rm r}(v, \tau) \Delta F_c(t-\tau)d\tau,
\label{eqn_rm_var}
\end{equation}
where $\Psi_{\rm e}(v, \tau)$ and $\Psi_{\rm r}(v, \tau)$ are called emissivity- and responsivity-weighted transfer functions, respectively, and
$\Psi_{\rm e}(v)$ is the delay integral of $\Psi_{\rm e}(v, \tau)$,
\begin{equation}
\Psi_{\rm e}(v) = \int \Psi_{\rm e}(v, \tau)d\tau.
\label{eqn_tf_emis_v}
\end{equation}
Similarly, the delay integrals of $\Psi_{\rm r}(v, \tau)$ is defined as
\begin{equation}
\Psi_{\rm r}(v) = \int \Psi_{\rm r}(v, \tau)d\tau.
\label{eqn_tf_resp_v}
\end{equation}
The centroid time delay can be calculated from the responsivity-weighted transfer function as
\begin{equation}
\tau_{\rm cent} = \frac{\iint \Psi_{\rm r}(v, \tau) \tau d\tau dv}{\iint \Psi_{\rm r}(v, \tau) d\tau dv}. 
\end{equation}
The velocity-resolved centroid time delays are calculated as
\begin{equation}
\tau_{\rm cent}(v) = \frac{\int \Psi_{\rm r}(v, \tau) \tau d\tau}{\int \Psi_{\rm r}(v, \tau) d\tau}.
\end{equation}

All the above formulae are model independent and apply to any sort of variable systems. Below, we will present specific expressions for them based on a simplified BLR photoionization model.
We will also illustrate that Equations~(\ref{eqn_rm_mean}) and (\ref{eqn_rm_var}) are a natural consequence of BLR reverberation.

\begin{figure*}
\centering
\includegraphics[width=1.0\textwidth]{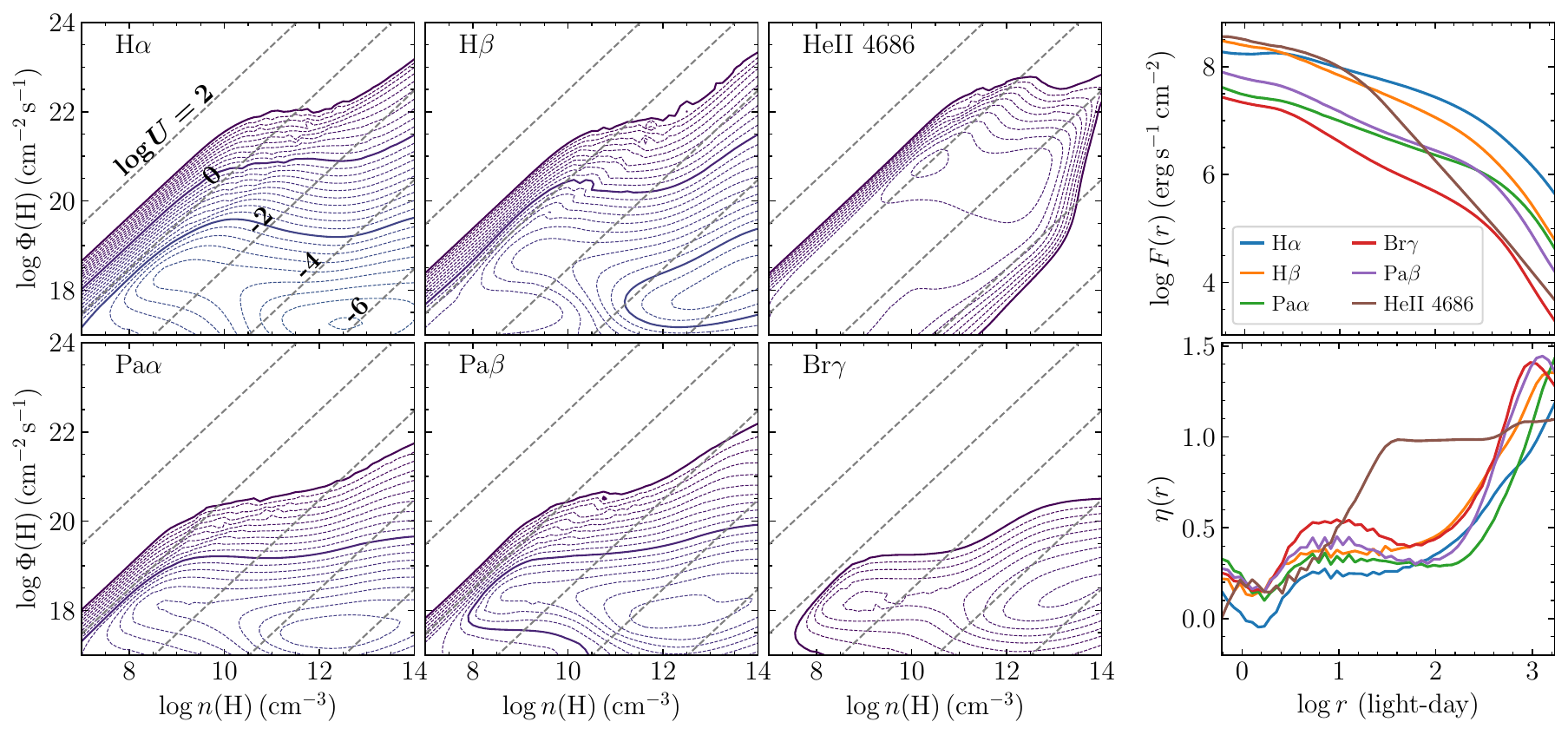}
\caption{The three left columns show contours of equivalent widths at 0.1 dex intervals for six emission lines (H$\alpha$, H$\beta$, \ion{He}{2} $\lambda4686$, Pa$\alpha$, Pa$\beta$, and Br$\gamma$) as a function of hydrogen number density $\log n(\rm H)$ and hydrogen-ionizing photon flux $\log \Phi(\rm H)$. The equivalent widths are referenced to the incident continuum at 1215~{\AA}. The smallest contour corresponds to 1~{\AA}, each solid line represents one dex, and dashed lines represent a step of 0.1 dex. The upper left blank region represents equivalent widths smaller than 1~{\AA}.
The grey dashed diagonal lines represent logarithm of the photoionization parameters ($\log U$) decreasing from 2 at upper left corner to -6 at lower right corner. The rightmost column shows the surface flux and responsivity with radius. The AGN spectral energy distribution is set by the default shape in \textsc{CLOUDY} code and the luminosity is set to $10^{42}~\rm erg~s^{-1}$ at 1350~{\AA}. The total hydrogen column density is $10^{23}~\rm cm^{-2}$.}
\label{fig_responsivity}
\end{figure*}

\subsection{Reverberation of BLRs}\label{sec_rm}
A basic consequence of BLR photoionization theory is that the line emissivity has a functional dependence on the incident continuum flux, which can be determined from photoionization calculations (e.g., \citealt{Goad2014}). For simplicity, we assume that
the emissivity of a single BLR cloud at a location $\bm{r}$ and time $t$ scales with a power of the incident continuum $f_c(\bm{r},t)$ as
\begin{equation}
\epsilon(\bm{r}, t) \propto f_c^{\eta}(\bm{r}, t),
\label{eqn_resp0}
\end{equation}
where $\eta$ is the non-linear response parameter. We assume that the continuum emission is isotropic so that the observed continuum is a good surrogate for the continuum illuminating the cloud. This is a quite reasonable assumption provided the BLR is not changing significantly during the period of RM monitoring. As a result, the incident continuum flux $f_c(\bm{r}, t)$ can be related to the observed continuum flux $F_c(t)$ as
\begin{equation}
f_c(\bm{r}, t) \propto \frac{D^2}{r^2} F_c(t-\tau_{\rm cl}),
\label{eqn_finc}
\end{equation}
where $D$ is the distance of the AGN from the observer, $r$ is the modulus of the vector $\bm{r}$, and $\tau_{\rm cl}$ is the time delay of the cloud depending on the location $\bm{r}$ (see below). By combining Equations~(\ref{eqn_resp0}) and (\ref{eqn_finc}), we can write the emissivity as
\begin{equation}
\epsilon(\bm{r}, t) = \frac{A(\bm{r})}{r^{2\eta}}F_c(t-\tau_{\rm cl}),
\label{eqn_resp1}
\end{equation}
where $A(\bm{r})$ is a factor including $D^{2\eta}$ and the proportionality factors implied in Equations~(\ref{eqn_resp0}) and (\ref{eqn_finc}). Again, the former can be determined from photoionization calculations.
We stress that here, the observed continuum flux should fully come from the ionizing source. In practice, the continuum flux is usually subject to contaminations from other components (e.g., the host galaxy starlight), however, those components can be separated out through certain methods, such as multi-component spectral and/or image decomposition.

From the above equation, it is straightforward to write
\begin{equation}
\frac{d\ln \epsilon}{d\ln F_c} = \eta.
\label{eqn_resp2}
\end{equation}
Here, $\eta$ is also termed {\it responsivity} in the literature (e.g., \citealt{Krolik1991, Goad1993, Korista2004}). Generally speaking, $\eta$ is not a constant and might vary with local physical conditions according to photoionization calculations shown in Figure~\ref{fig_responsivity} (see below).

Given a number distribution $f(\bm{r},\bm{w})$ of BLR clouds at a location $\bm{r}$ and velocity $\bm{w}$, the observed emission-line flux is obtained by summing over emissivity of all clouds expressed in Equation~(\ref{eqn_resp1}), namely,
\begin{equation}
F_l(t, v) 
= \iint \frac{A(\bm{r}) }{r^{2\eta}} F_c^{\eta}(t-\tau_{\rm cl})\delta(v-v_{\rm cl})f(\bm{r}, \bm{w}) d\bm{r}d\bm{w},
\label{eqn_rm_org}
\end{equation}
where $\delta$ is the Dirac delta function and the time delay $\tau_{\rm cl}$ of the cloud reads
\begin{equation}
\tau_{\rm cl} = \frac{r-\bm{r}\cdot\bm{n}}{c},
\label{eqn_tauc}
\end{equation}
where $c$ is the speed of light. The observed velocity of the cloud reads
\begin{equation}
v_{\rm cl} = -\bm{w}\cdot\bm{n},
\label{eqn_vc}
\end{equation}
where $\bm{n}$ denotes the unit vector of sightline pointing from the source to the observer and the appearance of a negative sign is because the velocity is defined in a way that a positive velocity means the cloud is receding from the observer. For an emission line with a rest-frame wavelength of $\lambda_0$, the velocity $v$ ($\ll c$) is related to wavelength via
\begin{equation}
\frac{v}{c} = \frac{\lambda-\lambda_0}{\lambda_0}.
\end{equation}

Under the condition that typical variation amplitudes of AGNs lie at a level of $\Delta F_c(t)/\bar F_c < 1$ (e.g., \citealt{Kelly2009, Lu2019}),
after some mathematical manipulation, it is easy to show that the emissivity- and responsivity-weighted transfer functions can be written as (see Appendix~\ref{sec_app_rm} for detailed derivations)
\begin{equation}
\Psi_{\rm e}(\tau, v) = \iint  \frac{A(\bm{r})}{r^{2\eta}} \bar F_c^{\eta-1}\delta(\tau-\tau_{\rm cl})\delta(v-v_{\rm cl})
f(\bm{r}, \bm{w})d\bm{r}d\bm{w}.
\label{eqn_tf_emis}
\end{equation}
and
\begin{equation}
\Psi_{\rm r}(\tau, v) = \iint  \eta \frac{A(\bm{r})}{r^{2\eta}} \bar F_c^{\eta-1} \delta(\tau-\tau_{\rm cl})\delta(v-v_{\rm cl})
f(\bm{r}, \bm{w})d\bm{r}d\bm{w},
\label{eqn_tf_resp}
\end{equation}
As can be seen, compared to $\Psi_{\rm e}$, there appears an extra factor $\eta$ in the integral for $\Psi_{\rm r}$, just reflecting that variations of the emission line are closely linked to the BLR responsivity.

From Equations~(\ref{eqn_tf_emis}) and (\ref{eqn_tf_resp}), we can respectively define the emissivity-weighted size as
\begin{equation}
R_{\rm EW} = \frac{\iint \bar F_c^{\eta}A(\bm{r})r^{-2\eta} f(\bm{r},\bm{w}) r d\bm{r}d\bm{w}}{\iint \bar F_c^{\eta}A(\bm{r})r^{-2\eta} f(\bm{r},\bm{w}) d\bm{r}d\bm{w}},
\end{equation}
and the responsivity-weighted size as (\citealt{Goad1993})
\begin{equation}
R_{\rm RW} = \frac{\iint \eta \bar F_c^{\eta}A(\bm{r})r^{-2\eta} f(\bm{r},\bm{w}) r d\bm{r}d\bm{w}}{\iint \eta \bar F_c^{\eta}A(\bm{r})r^{-2\eta} f(\bm{r}, \bm{w}) d\bm{r}\bm{w}}.
\label{eqn_r_rw}
\end{equation}
Combining Equations~(\ref{eqn_tauc}), (\ref{eqn_tf_resp}) and (\ref{eqn_r_rw}) immediately leads to a corollary that in general, for an extended BLR, $R_{\rm RW}$ is not equal to $c\tau_{\rm cent}$, unless the BLR is axi-symmetric and its emissions have no azimuthal structure.

\begin{figure}
\centering
\includegraphics[width=0.45\textwidth, trim=66pt 30pt 53pt 20pt, clip]{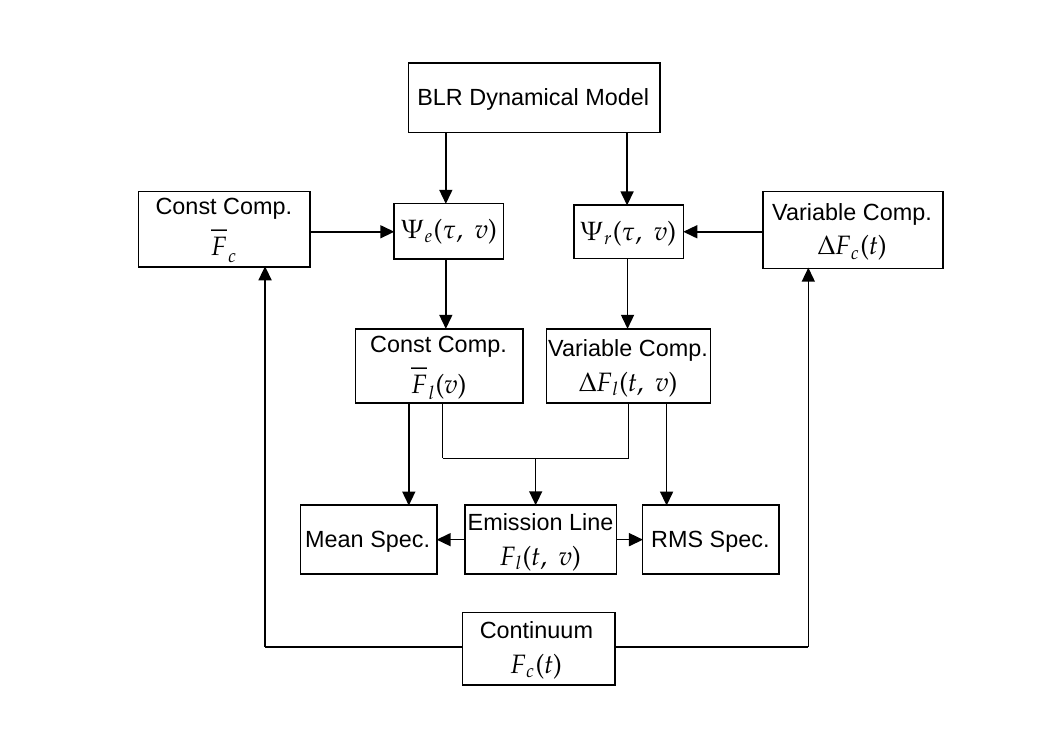}
\caption{A schematic diagram for calculating time series of emission lines given a BLR dynamical model with our framework. The descriptions of notations are summarized in Table~\ref{tab_variable}. }
\label{fig_flowchart}
\end{figure}

\subsection{RMS Spectra}\label{sec_rms}
By substituting  Equation~(\ref{eqn_rm_var}) into Equation~(\ref{eqn_rms_def}), one obtains that the RMS spectrum directly depends on the responsivity-weighted transfer function as (see Appendix~\ref{sec_app_rms} for a detailed derivation)
\begin{equation}
{\rm RMS} = \left[\iint \Psi_{\rm r}(\tau, v) \Psi_{\rm r}(\tau', v) S_c(\tau-\tau') d\tau d\tau'\right]^{1/2},
\label{eqn_rms}
\end{equation}
where $S_c(\tau-\tau')$ is the covariance function of the continuum variations at a time difference ($\tau-\tau'$). This equation implies that
the continuum variations also affect the RMS spectrum.

If the characteristic timescale of the continuum variations is much longer than the typical time delay of the transfer function, $S_c$
can be regarded as a constant in Equation~(\ref{eqn_rms}) and thereby the RMS spectrum is simply
\begin{equation}
{\rm RMS}  \propto \Psi_{\rm r}(v),
\label{eqn_rms1}
\end{equation}
where $\Psi_{\rm r}(v)\geqslant0$ is assumed.
That is to say, the RMS spectrum has exactly the same shape as the delay integral of the responsivity-weighted transfer function.
On the other hand, if the continuum variations are rapid so that the covariance function can be regarded as a delta function $S(\Delta \tau)\propto\delta(\Delta \tau)$, there will be
\begin{equation}
{\rm RMS} \propto \left[\int \Psi_{\rm r}^2(\tau, v) d\tau\right]^{1/2}\equiv \widehat{\Psi}_{\rm r}(v).
\label{eqn_rms2}
\end{equation}

For positive responsivity ($\eta>0$),  it is easy to find that the inequality $\Psi_{\rm r}(v)\geqslant\widehat{\Psi}_{\rm r}(v)$ is always satisfied according to the Cauchy–Schwarz inequality (\citealt{Steele2004}). At a given velocity bin, a more extended time delay distribution leads to an even larger $\Psi_{\rm r}(v)$ compared to $\widehat{\Psi}_{\rm r}(v)$. In realistic observations for a virialized BLR, the emission line core usually has a more extended time delay distribution. The distribution rapidly goes narrow towards line wings. This fact effectively results in a broader profile of $\widehat{\Psi}_{\rm r}(v)$ than that of $\Psi_{\rm r}(v)$. The underlying physical explanation is as follows. The outer region of the BLR with long time delays will smear out the rapid continuum variations and does not show effective responses. As a result, the emission line variations are mainly contributed from the inner region and thereby the RMS spectrum becomes broader. Indeed, the measured time delay tends to be shortened in case of rapid continuum variations. Such an effect had been explored and dubbed ``geometric dilution`` by \cite{Goad2014}. Here we demonstrate that ``geometric dilution`` can also affect the measured line widths of RMS spectra.

In generic cases, the RMS spectrum falls between the above two extreme cases, so that it will be somehow different from either $\Psi_{\rm r}(v)$ or $\widehat{\Psi}_{\rm r}(v)$, depending on the timescale of continuum variations.
It has been established from RM campaigns that the typical time delay of H$\beta$ BLRs is tightly correlated with 5100~{\AA} luminosity as $\tau_{\rm H\beta}=33 l_{44}^{0.533}$ days, where $l_{44}=L_{\rm 5100}/10^{44}~\rm erg~s^{-1}$ (see \citealt{Bentz2013, Du2019}). Using the damped random walk process to model AGN light curves, previous studies found that typical variation timescale is also correlated with 5100~{\AA} luminosity as $\tau_{\rm d}=52 l_{44}^{0.46}$ days, although with a quite large scatter (\citealt{Li2013, Lu2019}). This implies that AGN variation timescales are comparable to typical H$\beta$ time lags and the effect of geometric dilution on line widths might be not negligible.

Below we elaborate on the resulting RMS spectra in cases of constant and radial-dependent responsivity separately.
\paragraph{Constant responsivity.}
If the responsivity is a constant spatially and temporally, we can rewrite Equation~(\ref{eqn_rm_org}) in a concise convolutional form involving the whole continuum and emission line fluxes
\begin{equation}\label{eqn_rm_const}
F_l(t, v) = \int \Psi_{c}(\tau, v) F_c^{\eta}(t-\tau) d\tau,
\end{equation}
where the transfer function $\Psi_c$ is related to above defined $\Psi_{\rm e}$ or $\Psi_{\rm r}$ by
\begin{equation}
\Psi_c(\tau, v) = \bar F_c^{1-\eta} \Psi_{\rm e}(\tau, v) = \eta^{-1}\bar F_c^{1-\eta} \Psi_{\rm r}(\tau, v).
\end{equation}
Meanwhile, the mean spectrum is exactly identical in shape to the delay integral of the transfer functions,
\begin{equation}
 \bar F_l(v) \propto \Psi_{\rm r}(v)\propto \Psi_{\rm e}(v).
\end{equation}
If the variation timescale of the continuum is much longer than the typical time delay of the BLR,
the mean and RMS spectra will also be exactly identical in shape  (see Equation~\ref{eqn_rms1}).

\begin{deluxetable*}{ccl}
\tablecolumns{3}
\tabletypesize{\footnotesize}
\tabcaption{\centering Fiducial Parameters of the BLR Model. \label{tab_parameter}}
\tablehead{
\colhead{Parameter} & \colhead{Value} & \colhead{Description}
}
\startdata
$R_{\rm BLR}$       &  10 light-days & Mean radius\\
$\beta$             &  1.0           & Shape parameter of the radial distribution\\
$F_{\rm in}$        &  0.2           & Inner edge in units of $R_{\rm BLR}$ \\
$\theta_{\rm inc}$  & 30$^\circ$     & Inclination angle \\
$\theta_{\rm opn}$  & 30$^\circ$     & Openning angle \\
$M_\bullet$         & $10^7M_\odot$  & Black hole mass \\
$\eta_0$            & 0.2            & Constant coefficient of the responsivity $\eta(r)$ in Equation~(\ref{eqn_eta})\\
$\eta_1$            & 0.2            & Power law amplitude of the responsivity $\eta(r)$ in Equation~(\ref{eqn_eta})\\
$\alpha$            & 2.0            & Power law index of the responsivity $\eta(r)$ in Equation~(\ref{eqn_eta})\\\hline
$R_{\rm RW}$        & 15.7 light-days & Responsivity-weighted radius\\
$R_{\rm EW}$        & 10   light-days & Emissivity-weighted radius
\enddata
\end{deluxetable*}

\begin{figure*}[thp!]
\centering
\includegraphics[width=0.9\textwidth]{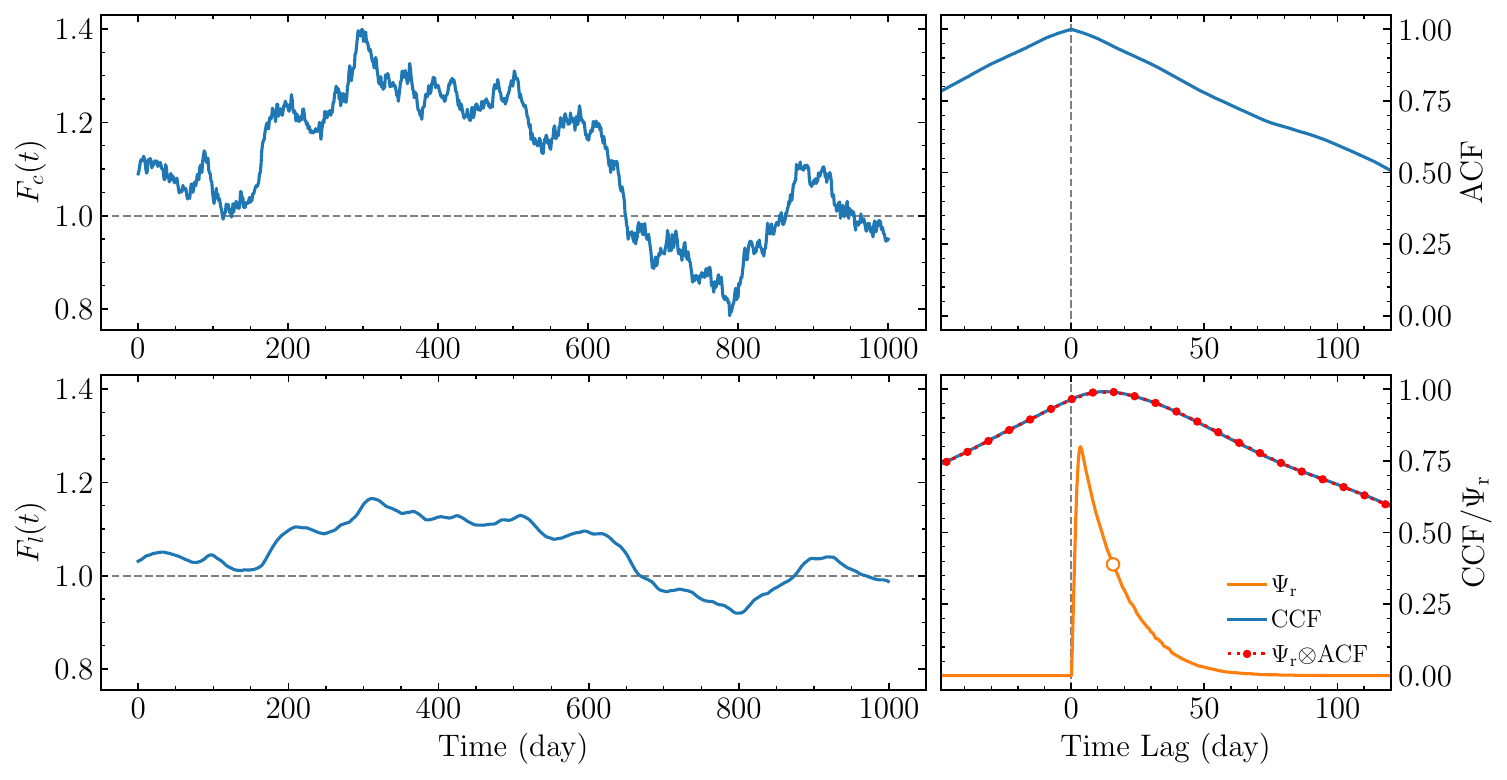}
\caption{An illustrative simulation to show the generated continuum and emission line light curves in left panels and the ACF of the continuum light curve and CCF between the continuum and emission line light curves. In the left panels, the horizontal dashed lines represent $\bar F_{\rm c}$ and $\bar F_l$, respectively. In the lower right panel, the yellow line plots the responsivity-weighted transfer function $\Psi_{\rm r}(\tau)$ with the circle point marking the centroid time lag. The red dotted line with solid points plots the convolution between $\Psi_{\rm r}(\tau)$ and ACF of the continuum light curve, which is exactly coincident with the CCF.}
\label{fig_lc}
\end{figure*}

\paragraph{Radial-dependent responsivity.}
If the responsivity is a function of radius, the resulting mean and RMS spectra are usually different in shape. For virialized BLRs, the appearance of an extra factor $\eta$ in the integral of $\Psi_{\rm r}$ (compared to $\Psi_{\rm e}$; see Equations~\ref{eqn_tf_emis} and \ref{eqn_tf_resp}) causes the mean spectrum to be broader than the RMS spectrum for radially increasing responsivity. Accordingly, the opposite cases of radially decreasing responsivity will cause the RMS spectrum  to be broader.

To illustrate how responsivity changes with radius, we calculate surface emissivity of six prominent broad emissions using the spectral synthesis code \textsc{CLOUDY} (version 23.01; \citealt{Gunasekera2023}) over a grid of hydrogen number density of BLR clouds and hydrogen-ionizing photon flux.
We adopt the default AGN spectral energy distribution (\citealt{Mathews1987}) in \textsc{CLOUDY} and set a luminosity of $10^{42}~\rm erg~s^{-1}$ at 1350~{\AA}, scaled for a $10^7M_\odot$ SMBH according to the observation of NGC~5548 (\citealt{Korista2004}). The total hydrogen column density is fixed to $10^{23}~\rm cm^{-2}$. In Figure~\ref{fig_responsivity}, we show the obtained responsivity with radius for different emission lines. As can be seen, the responsivity is different for different emission lines but overall increases with radius. In addition, the responsivity deviates from $\eta=1$, except for \ion{He}{2}~$\lambda4686$ at large radius. This illustrates that most emission lines respond non-linearly to continuum variations.

\begin{figure*}[th!]
\centering
\includegraphics[width=0.95\textwidth]{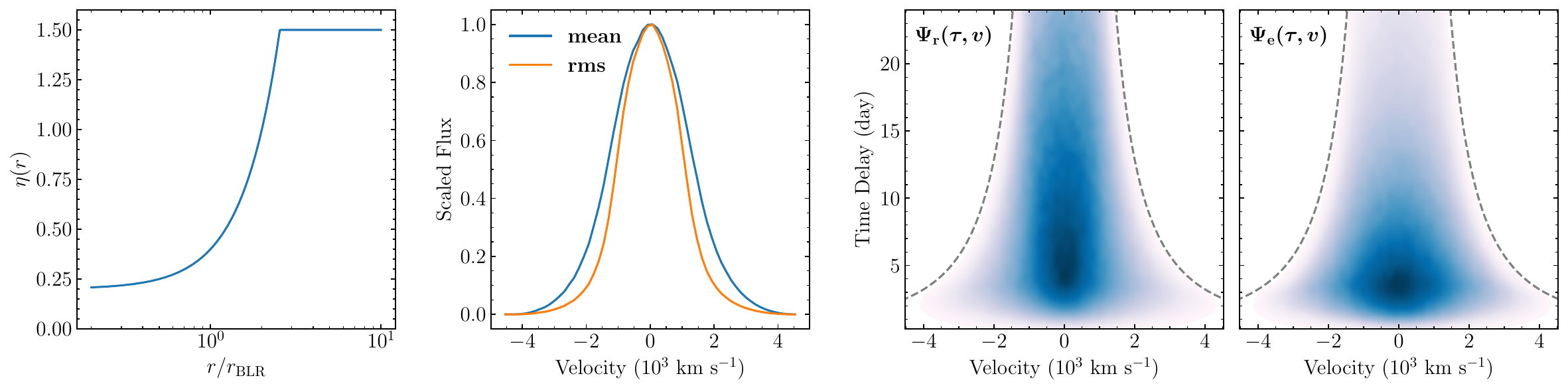}
\includegraphics[width=0.95\textwidth]{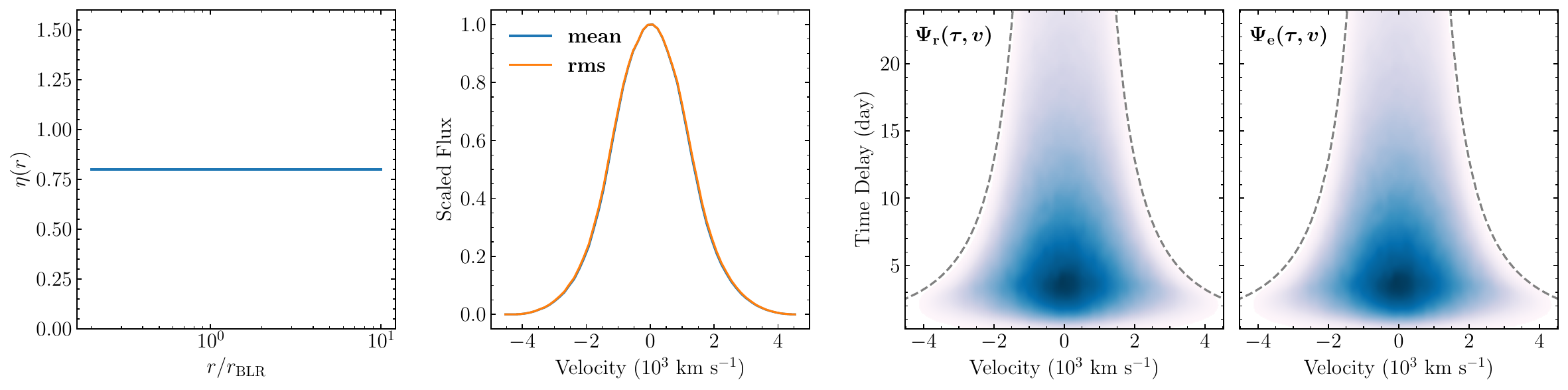}
\includegraphics[width=0.95\textwidth]{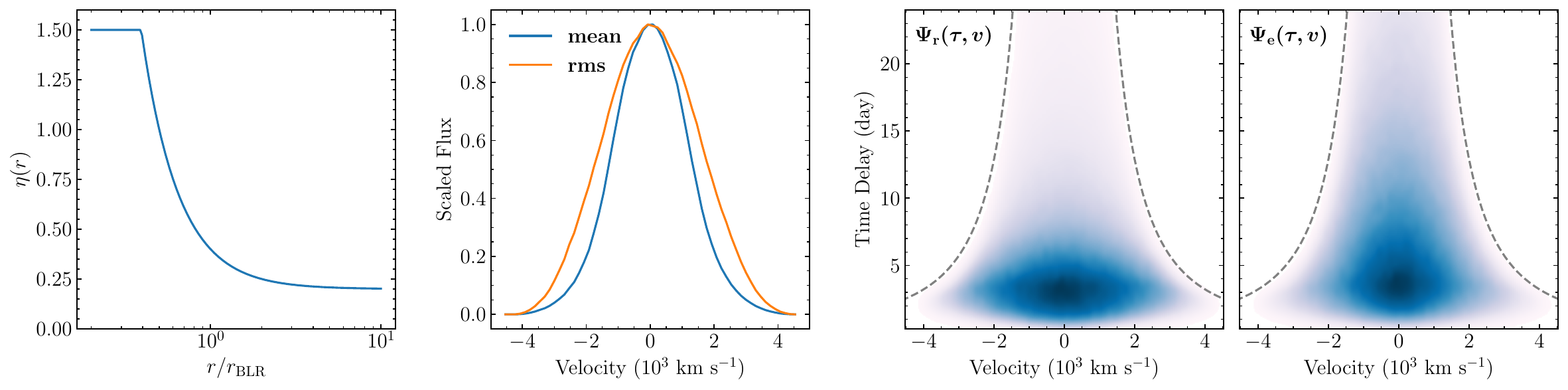}
\caption{From left to right panels are the adopted form of responsivity as a function of radius $\eta(r)$, the resulting mean and rms spectra, the responsivity-weighted transfer function $\Psi_{\rm r}(\tau, v)$, and the emissivity-weighted transfer function $\Psi_{\rm e}(\tau, v)$, respectively. The grey dashed lines represent the virial envelope ($v^2=GM_\bullet/c\tau$, where $M_\bullet=10^7M_\odot$). }
\label{fig_mean_rms}
\end{figure*}

\section{Exemplary Simulations}\label{sec_example}
\subsection{A Toy BLR Dynamical Model}
In this section, we construct a simple parameterized BLR model to demonstrate the effects of responsivity. From the viewpoint of dynamical modeling, in the right-hand side of Equation~(\ref{eqn_rm_org}), the terms $A(\bm{r})r^{-2\eta}$ and $f(\bm{r}, \bm{w})$ are degenerate and largely unknown. In RM campaigns, the mean spectrum and variations of an emission line are directly observable. Equations~(\ref{eqn_mean}) and (\ref{eqn_var}) in Appendix illustrates that the mean spectrum puts constraints on the overall product $A(\bm{r})r^{-2\eta}f(\bm{r}, \bm{w})$, whereas variations of the emission line put extra constraints on the responsivity $\eta$. Inspired by this factor, we parameterize the responsivity and other terms separately, which greatly simplifies our dynamical modeling. Such a treatment also enables us to concentrate on exploring responsivity rather than delving into the detailed physical conditions of BLRs. In this sense, our treatment is idealized but adequate for illustration purpose.

For the responsivity, we parameterize it with a power law as
\begin{equation}
\eta(r) = \eta_0 + \eta_1 \left(\frac{r}{r_{\rm BLR}}\right)^\alpha,
\label{eqn_eta}
\end{equation}
where $r_{\rm BLR}$ is a characteristic radius of the BLR and $\eta_0$, $\eta_1$, and $\alpha$ are free parameters. To align with photoionization calculations shown in Figure~\ref{fig_responsivity}, we further limit the responsivity to a range
\begin{equation}
-0.5\leq \eta(r) \leq 1.5.
\end{equation}
A negative $\eta$ is possible for rarefied BLRs with a low gas density, plausibly manifesting as anomalous responses of emission lines (e.g., \citealt{Du2023}).
In our calculations, those $\eta$ outside the above range are forced to take the lower/upper limits.

To model the product $A(\bm{r})r^{-2\eta}f(\bm{r}, \bm{w})$, we simply assume the number distribution $f(\bm{r}, \bm{w})$ depends only on the location $\bm{r}$ and the cloud's velocity is fully determined given $\bm{r}$ (such as Keplerian rotation). This means that we can simplify $f(\bm{r}, \bm{w})$ into a distribution as a function of only $\bm{r}$, namely, $f(\bm{r}, \bm{w})\equiv g(\bm{r})$. The cloud distribution is assumed to be axi-symmetric and uniform over $\cos\theta$ in the $\theta$-direction. As such, we only need to specify the radial distribution $g(r)$. Also, $A(\bm{r})$ depends only on $r$. We introduce an overall parameterization prescription using the Gamma distribution (e.g., \citealt{Pancoast2011})
\begin{equation}
A(r)r^{-2\eta}g(r) \propto \left(\frac{r}{s}\right)^{k-1}\exp\left( -\frac{r}{s}\right),
\label{eqn_gam}
\end{equation}
where $s$ is a scale parameter controlling the spatial extent and $k$  is a shape parameter controlling the radial profile. This distribution was commonly used in previous BLR dynamical modeling (\citealt{Pancoast2011, Pancoast2014, Li2013, Li2018, Li2022}). It has flexibility in approximating exponential ($k=1$),  single-peaked (with a large $k$), or long-tailed (with a small $k$) distributions. There are also other forms of distributions, such as a power law, used in the literature (\citealt{Stern2015, Li2022}). Nevertheless, the distribution in Equation~(\ref{eqn_gam}) is sufficient for present illustration purpose. To further simplify our parameterization, we define (\citealt{Pancoast2011})
\begin{equation}
R_{\rm BLR} = ks, ~~~~\beta = 1/\sqrt{k}.
\end{equation}
The distribution is then characterized by a mean of $R_{\rm BLR}$ and standard deviation of $\beta R_{\rm BLR}$, which implies that $\beta$ controls the spatial extension of BLR clouds.

The cloud emission is isotropic and self-shadowing effect is neglected for simplicity. Each cloud orbits around the central SMBH with circular Keplerian motion. All clouds constitute a thick-disk like geometry, with an openning angle of $\theta_{\rm opn}$ and inclination angle of $\theta_{\rm inc}$ (see Figure~1 of \citealt{Li2013} for the meaning of these two parameters). Table~\ref{tab_parameter} summarizes the model parameters and their adopted values, together with the resulting emissivity- and responsivity-weighted BLR sizes. Figure~\ref{fig_flowchart} shows a schematic diagram for generating simulated time series of emission lines using the framework described in Section~\ref{sec_form}. We have incorporated the present framework into our previously developed dynamical modeling code \textsc{BRAINS}\footnote{The living code of \textsc{BRAINS} is publicly available at \url{https://github.com/LiyrAstroph/BRAINS}, while the version used in this work is available at \url{https://doi.org/10.5281/zenodo.10674858}.} (\citealt{Li2013, Li2018}). Below we employ \textsc{BRAINS} to perform simulations. It is worth iterating that since we parameterize $\eta$ and other terms in Equation~(\ref{eqn_rm_org}) separately, a change in responsivity affects only the variable component (and hence RMS spectrum) of the emission line, but does not affect the mean spectrum.

\begin{figure*}[ht!]
\centering
\includegraphics[width=0.85\textwidth]{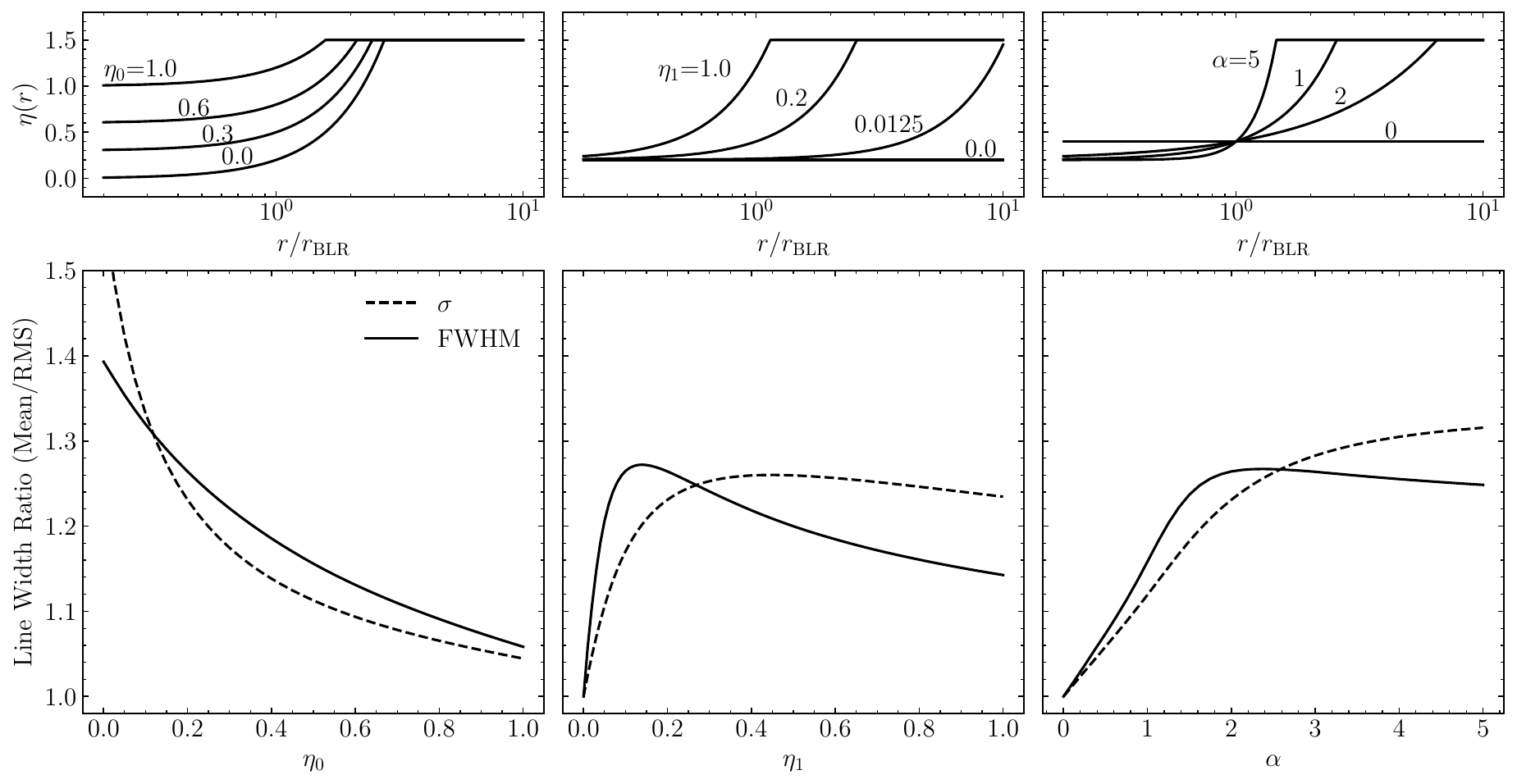}
\caption{Bottom panels show the width ratios between mean and RMS spectra for different values of (left) $\eta_0$, (middle) $\eta_1$, and (right) $\alpha$.
Solid and dashed lines are for FWHM and line dispersion $\sigma$, respectively. Top panels show the corresponding forms of $\eta(r)$. The fiducial values are $\eta_0$=0.2, $\eta_1$=0.2, and $\alpha=2$.}
\label{fig_width}
\end{figure*}
\begin{figure*}
\centering
\includegraphics[width=0.7\textwidth]{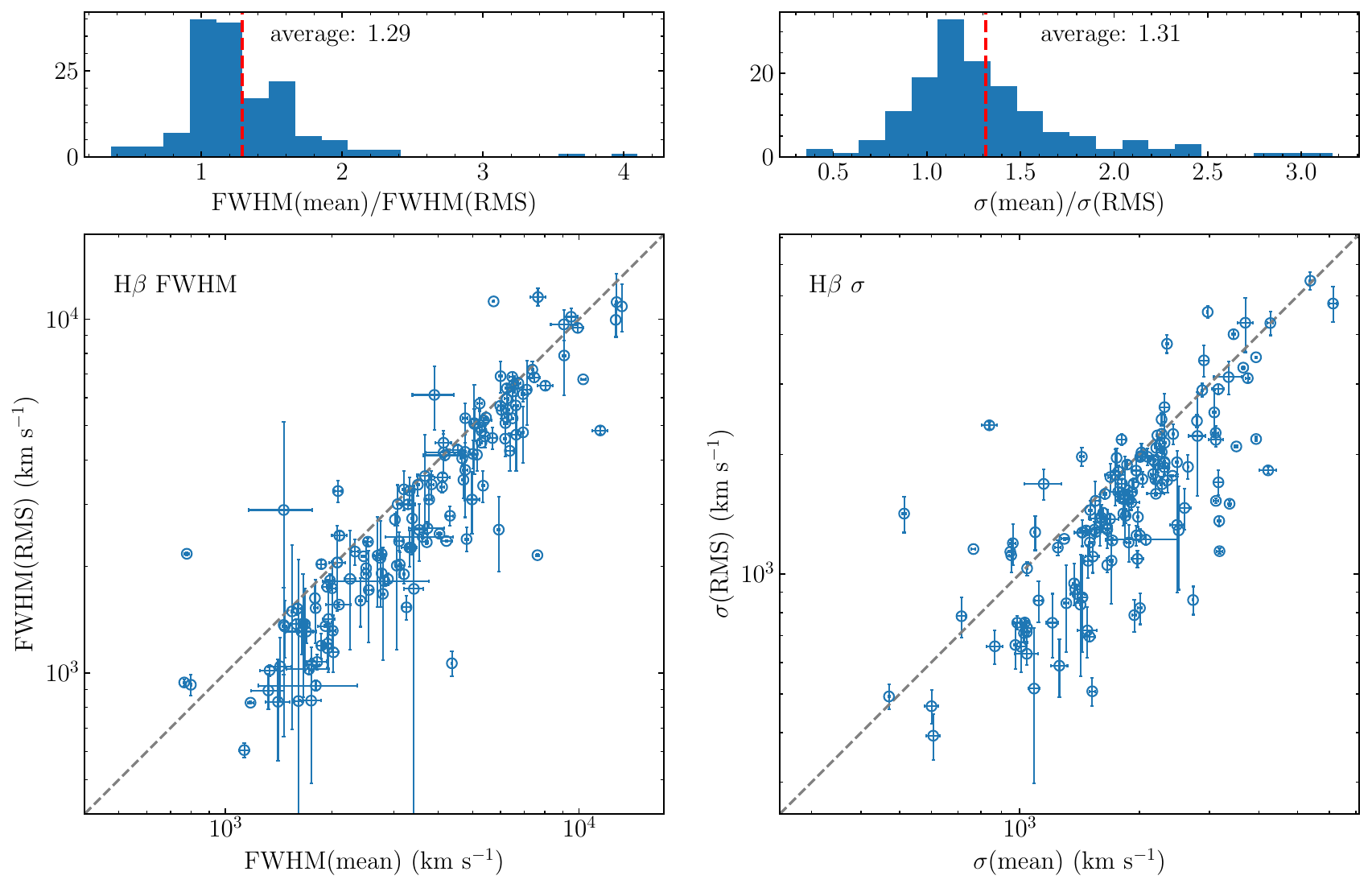}
\caption{Comparisons between H$\beta$ line widths of mean and RMS spectra compiled from previous RM campaigns in bottom panels, together with histograms of the corresponding width ratios in top panels.  The left and right panels are for the FWHM and line dispersion, respectively. The red vertical dashed lines represent the averaged width ratios.}
\label{fig_width_obs}
\end{figure*}

\begin{figure*}[!ht]
\centering
\includegraphics[width=0.85\textwidth]{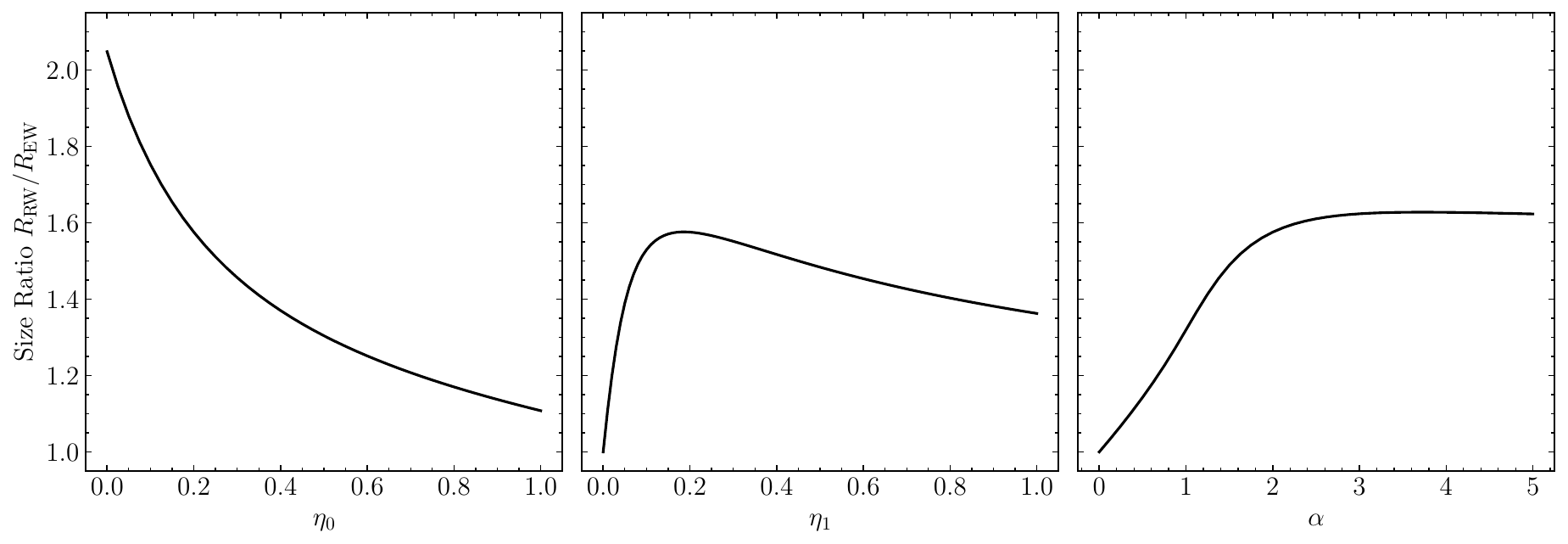}
\caption{The ratio between responsivity- and emissivity-weighted BLR radius ($R_{\rm RW}/R_{\rm EW}$) for different values of $\eta_0$, $\eta_1$, and $\alpha$.
The adopted form of $\eta(r)$ are the same as those in the upper panels of Figure~\ref{fig_width}. The fiducial values are $\eta_0$=0.2, $\eta_1$=0.2, and $\alpha=2$.}
\label{fig_size}
\end{figure*}

\subsection{An Example of Simulated Light Curves}
Using the dynamical model described above, we showcase a pair of simulated continuum and emission line light curves in Figure~\ref{fig_lc}. The continuum variability follows the damped random walk process (e.g., \citealt{Kelly2009, Li2014}) with a damping timescale of $\tau_{\rm d}=200$ days, much longer than the input typical time delay ($\sim$15.7 days). Therefore, the geometric dilution effect on the line widths in the RMS spectrum is not important. As can be seen, the continuum varies with a large amplitude of $\sim$30\%, whereas the emission line varies quite mildly, only with an amplitude less than 15\%. Besides variation blurring due to the spatial extension of the BLR, the adopted low responsivity ($\eta_0=0.2$) further reduces variability of the emission line.

We employ the interpolated cross-correlation method (\citealt{Gaskell1987}) to calculate the auto correlation function (ACF) of the continuum light curve and the cross-correlation function (CCF) between the continuum and emission line, plotted in the right panels of Figure~\ref{fig_lc}. It is known that the CCF is indeed equivalent to the convolution between the transfer function and ACF of the continuum (e.g., \citealt{Welsh1999, Li2013}). The lower right panel of Figure~\ref{fig_lc} also superimposes the responsivity-weighted transfer function and its convolution with the ACF, which is in good agreement with the CCF as expected.
The responsivity-weighted transfer function has a peak time lag of 3.4 days and a centroid time lag of 15.7 days. The latter is exactly equal to the responsivity-weighted radius listed in Table~\ref{tab_parameter} since the BLR model is axi-symmetric and the emissions are isotropic.  The CCF yields a peak time lag of 12.6 days and a centroid time lag of 15.1 days. This centroid value slightly differs from that of the transfer function because of the influences of the ACF's shape as well as the adopted criterion for the CCF points used to calculate the centroid. Following the convention, we use the CCF points above 80\% of the peak value (e.g., \citealt{Peterson1999}). A different criterion might slightly change the centroid time lag (\citealt{Koratkar1991}).

\subsection{Mean and RMS Spectra}
In Figure~\ref{fig_mean_rms}, we show the resulting mean and RMS spectra, responsivity- and emissivity-weighted transfer functions for different forms of responsivity. As expected, in the case of $\eta$ increasing with radius, the responsivity-weighted transfer function is more extended along the time delay direction compared to emissivity-weighted transfer function, implying that there is significant responsivity of the emission line at large radius. The resulting RMS spectrum is therefore narrower than the mean spectrum. The results are just the opposite for $\eta$ decreasing with radius. The middle row of Figure~\ref{fig_mean_rms} also confirms that for the case of spatially constant responsivity, the mean and RMS spectra are exactly identical, even when $\eta$ deviates from one (corresponding to linear response).

In Figure~\ref{fig_width}, we further illustrate how the ratio of line widths from mean and RMS spectra changes with the parameters $\eta_0$, $\eta_1$, and $\alpha$ of the responsivity (see Equation~\ref{eqn_eta}). As mentioned above, a spatially constant responsivity leads to an identical width ratio of mean and RMS spectra. The  converse also applies, namely, a large radial change of responsivity results in a large difference in the width ratio. In the fiducial case of $\eta_1=0.2$ and $\alpha=2$, the width ratio can be up to about 1.5 for $\eta_0=0$. In the middle panels of Figure~\ref{fig_width}, there appears a maximum value of width ratio around $\eta_1\sim0.1$ for FWHM and $\sim0.4$ for line dispersion. This is because as $\eta_1$ increases beyond these critical values, the responsivity approaches the upper limit of 1.5 for most radii, in turn narrowing down the region with significant changes of responsivity. This dilutes the differences between mean and RMS spectra. The situation is similar for the dependence on $\alpha$. We note that the different dependence for FWHM and line dispersion boils down to the non-Gaussian line profiles, as illustrated in Figure~\ref{fig_mean_rms}.

\begin{figure*}[!ht]
\centering
\includegraphics[width=0.85\textwidth]{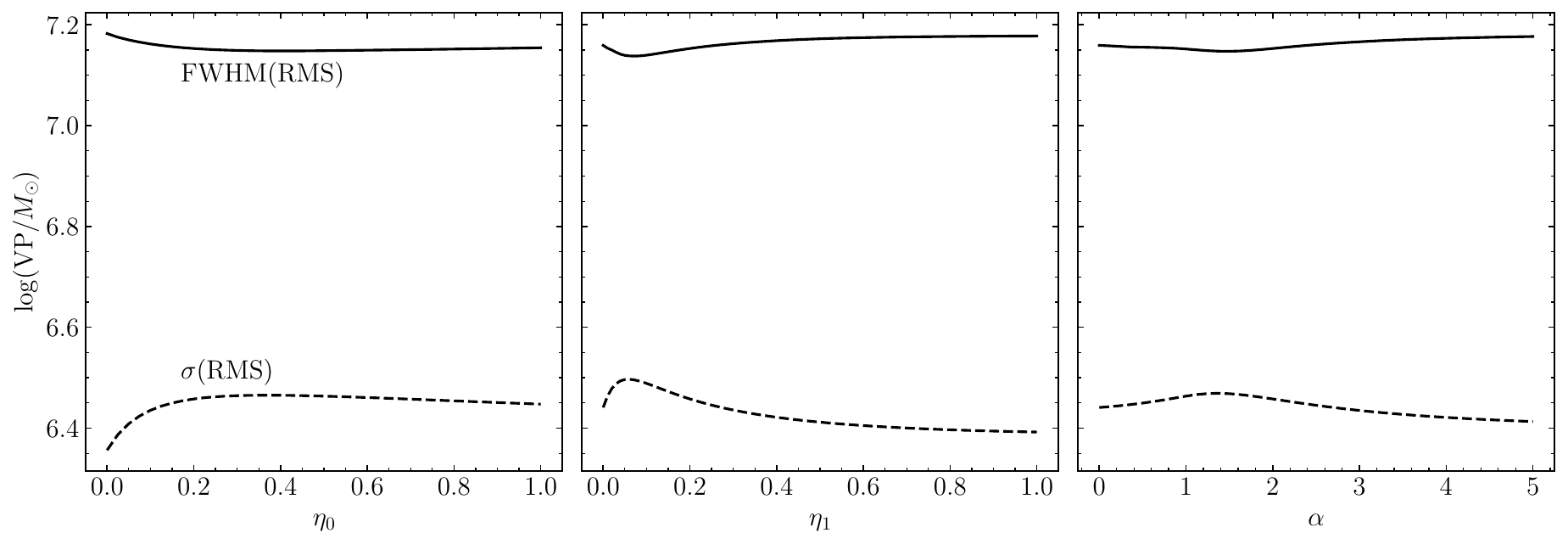}
\caption{The virial products calculated using FWHM and line dispersion of RMS spectra for different values of $\eta_0$, $\eta_1$, and $\alpha$.
The adopted form of $\eta(r)$ are the same as those in the upper panels of Figure~\ref{fig_width}. The fiducial values are $\eta_0$=0.2, $\eta_1$=0.2, and $\alpha=2$.}
\label{fig_mass}
\end{figure*}

\section{Observational Implications}\label{sec_implication}

\subsection{Line Widths from Mean and RMS Spectra}
In Figure~\ref{fig_width_obs}, we compile H$\beta$ line widths (FWHM and line dispersion) of mean and RMS spectra from previous RM campaigns. Most of the data are directly extracted from Table 3 of \cite{Yan2024}. We additionally include the latest measurements for a dozen of AGNs from \cite{Hu2024}, \cite{Li2024}, and \cite{Yao2024}. We can find that the majority of data points lie under the line of equality. The averaged ratio of FWHM between mean and RMS spectra is about 1.29, indicating that FWHM(mean) is systematically broader than FWHM(RMS) by $\sim$29\%. The averaged ratio of line dispersion has a similar value of about 1.31.
Combining with Figure~\ref{fig_width}, the systematically broader mean spectra of the H$\beta$ line point to a radial increasing responsivity, consistent with photoionization calculations in Figure~\ref{fig_responsivity}. With the present simple BLR model, these width ratios correspond to a rough range of $\eta_0\sim0.1-0.2$, $\eta_1\sim0.1-0.3$, and $\alpha\sim2-5$. Our analysis illustrates that the observed line width ratio of mean and RMS spectra provides a useful constraint on BLR photoionization models, which can be further tested with detailed photoionization calculations.


\subsection{BLR Sizes from Spectroastrometry and Reverberation Mapping}
As mentioned above, the SA technique with the GRAVITY/VLTI instrument provides a new diagnostic of BLRs, complementary to RM  (e.g., \citealt{Li2022, Li2023}). However, there are important differences in the measured BLR sizes between the two methods. Firstly, RM relies on time delay analysis and probes the BLR structure perpendicular to iso-delay paraboloids. As a comparison, SA invokes photocenter offsets of BLRs projected on the sky and probes the BLR structure perpendicular to the line of sight (e.g., \citealt{Beckers1982, Bailey1998}). These two methods are sensitive to different dimensions of BLRs. Secondly and more importantly, RM measures responsivity-weighted sizes, while SA observes instantaneous emissions from BLRs and thereby measures emissivity-weighted sizes. As illustrated in Section~\ref{sec_example}, once the responsivity is not a global constant, the responsivity- and emissivity-weighted sizes will be no longer equivalent. Figure~\ref{fig_size} shows that the difference between the two sizes can exceed 50\%, depending on how the responsivity varies with radius.

Such differences in emissivity- and responsivity-weighted BLR sizes are crucial for SARM analysis (\citealt{Wang2020, Li2022}), which offers a more thorough view of geometry and kinematics of BLRs and thus is expected to improve accuracy of SMBH mass measurements. In particular, SARM analysis leverages the fact that SA measures the angular size while RM measures the physical size of the BLR. Their combination directly yields a geometric distance to BLRs (\citealt{Elvis2002, Rakshit2015, Wang2020}). \cite{Zhang2021} conducted photoionization calculations for a simplified BLR configuration in which both gas density distribution and radial cover factor are parameterized by power laws. Their results also demonstrated systematic deviations between BLR sizes measured from SA and RM, confirming the factors described above. In this work, we establish a unified framework to delineate both emissivity- and responsivity-weighted BLR sizes, which enables a self-consistent, practically feasible BLR dynamical modeling approach for SARM analysis. This will largely alleviate possible systematics arising from previous SARM analysis that resorts to globally linear response of BLRs (\citealt{Wang2020, GRAVITY2021, Li2022}) and thereby improve the distance determinations.

\begin{figure*}
\centering
\includegraphics[width=0.7\textwidth]{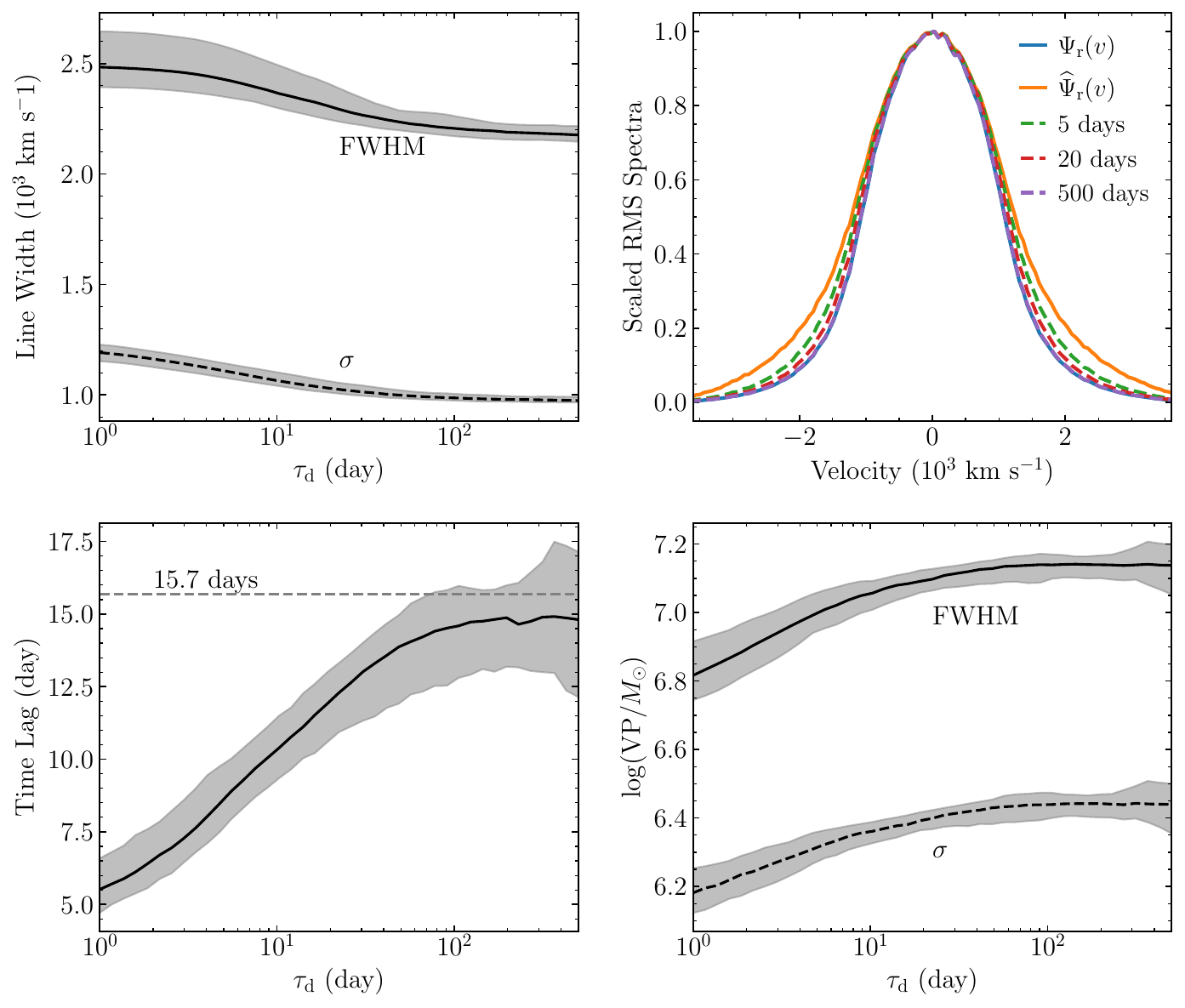}
\caption{(Upper left) the emission line widths of RMS spectra,  (bottom left) time delays measured using the interpolated CCF method, and (bottom right) virial products for different continuum variation timescale $\tau_{\rm d}$ from simulations with the BLR model parameters listed in Table~\ref{tab_parameter}. The solid lines represent the median values and grey shaded areas represent the 90\% confidence intervals.  The upper right panel shows examples of RMS spectra. Here, $\tau_{\rm d}$ is the damping timescale of the damped random walk process used to generate continuum light curves.}
\label{fig_taud}
\end{figure*}

\subsection{Black Hole Mass Measurements}
The commonly used recipe for measuring SMBH masses with RM is based on the virial product, namely,
\begin{equation}
M_\bullet = f_{\rm BLR} M_{\rm vp} = f_{\rm BLR}\frac{c\tau (\Delta V)^2}{G},
\label{eqn_mass}
\end{equation}
where $f_{\rm BLR}$ is the virial factor, $G$ is the gravitational constant, and $\Delta V$ is the line width.
The time delay $\tau$ is usually measured from the CCF method, which relies on variation components of the continuum and emission lines.
From our formulation of RM described in Section~\ref{sec_rm}, it is obvious that such a time delay should be directly associated with the responsivity-weighted (instead of emissivity-weighted) transfer function, which in turn controls the RMS spectrum of the emission line (see Equations~\ref{eqn_rms1} and \ref{eqn_rms2}). Therefore, an appropriate choice of line width measure should point to the RMS spectrum, rather than the mean spectrum. Figure~\ref{fig_width_obs} clearly demonstrates that the line widths measured from mean and RMS spectra of AGNs with RM observations are not the same and their ratios span broad distributions.
In the present framework, those broad distributions reflect that the responsivity might change from object to object.

In Figure~\ref{fig_mass}, we calculate virial products ($M_{\rm VP}$) for different values of $\eta_0$, $\eta_1$, and $\alpha$. To avoid any random noise from light curve realizations, we directly use the responsivity-weighted radius $R_{\rm RW}$ as a substitute to $c\tau$ in Equation~(\ref{eqn_mass}). It is interesting to note that the virial product is not a constant but slightly varies with these parameters within a level of $\sim$0.05 dex. It seems that the FWHM yields a more consistent virial product than the line dispersion. This is because the RMS spectrum of the emission line changes in shape with the responsivity parameters and the line dispersion is more sensitive to the line shape than the FWHM.

In Section~\ref{sec_rms}, we have proved that the RMS spectrum indeed depends on the variation timescale of the continuum light curve (see Equation~\ref{eqn_rms1} and \ref{eqn_rms2}) because of the ``geometric dilution'' effect. Here, we perform a more specific simulation to illustrate this effect. Using BLR model parameters listed in Table~\ref{tab_parameter}, we randomly generate a batch of mock continuum and emission line data, both with a duration of 1000 days, for different damping timescale $\tau_{\rm d}$ of the damped random walk process. We then calculate line widths (FWHM and line dispersion) of the RMS spectra, measure time delays using the interpolated CCF methods, and determine the virial products using Equation~(\ref{eqn_mass}). Figure~\ref{fig_taud} shows medians and 90\% confidence intervals of the obtained line widths, time delays, and virial products, as well as several selected examples of RMS spectra at $\tau_{\rm d}=5$, 20, and 500 days. As expected, the median line widths decrease with increasing $\tau_{\rm d}$, by $\sim$10\% for FWHM and  $\sim20$\% for line dispersion from $\tau_{\rm d}=1$ to 500 days. As illustrated in the bottom right panel of Figure~\ref{fig_lc}, the extension of the transfer function along the time delay direction spans no more than 50 days, therefore, the RMS spectrum approaches $\Psi_{\rm r}(v)$ and the resulting line widths tend to be stable when $\tau_{\rm d}\gtrsim50$ days.

Because of the ``geometric dilution'' effect, the resulting time delays also depend on the continuum variation timescale. This has been well explored by \cite{Goad2014}. In our simulations, the time delays suffer to a quite severe influence (compared to the line widths), increasing from $\sim$5 days to the expected $\sim$15.7 days ($=R_{\rm RW}$) for $\tau_{\rm d}$ from 1 to 500 days. As a result, the virial products are subject to a deviation as large as $\sim$0.3 dex at small $\tau_{\rm d}$. In realistic RM observations, the continuum variation timescale might not be such extremely short compared to the characteristic time delay. Nevertheless, the influences of continuum variation timescales need to be taken into account for accurate SMBH mass measurements.

\subsection{BLR Dynamical Modeling}
As we mentioned above, both photoionization calculations (Figure~\ref{fig_responsivity}) and RM observations (Figure~\ref{fig_width_obs}) imply a radial-dependent responsivity of BLRs. A direct consequence is that RM analysis needs to account for both constant and variable components of the emission line, as expressed in Equations~(\ref{eqn_fl}) and (\ref{eqn_rm_full}). However, to our knowledge, most of previous BLR dynamical modeling approaches (e.g., \citealt{Pancoast2011, Pancoast2014, Rosborough2024, Williams2022}) assumed a linear response of BLRs, effectively equivalent to $\eta(r)\equiv1$ in present framework. \cite{Li2013, Li2018} included a non-linear response but still forced the responsivity to be constant across the BLR (equivalent to using Equation~\ref{eqn_rm_const}). As a result, all previous BLR dynamical modeling approaches cannot appropriately cover the differences between mean and RMS spectra. This fact had been recognized by \cite{Mangham2019}, which employed a disk wind model to simulate time series of emission lines and found that the dynamical model code of \cite{Pancoast2014} failed to reproduce the RMS spectra. In addition, \cite{Pancoast2018} performed a comparison of transfer functions obtained from the dynamical modeling and the maximum entropy method (\citealt{Horne1994}) by applying these two methods on the H$\beta$ RM data of an individual AGN Arp~151 (\citealt{Bentz2009}). Their results showed that the two transfer functions seem not to be well matched with each other, although a quantitative assessment cannot be made because of the difficulty in reliable uncertainty estimates for the maximum entropy method. Indeed, the maximum entropy method solves the linearized RM equation, somehow equivalent to Equation~(\ref{eqn_rm_var}) in the present framework. We note that for the RM data of Arp 151 in 2008 season (\citealt{Bentz2009}), the H$\beta$ line widths measured from the mean spectrum are significantly broader than those from the RMS spectrum: FWHM(mean)/FWHM(RMS)$\sim$$1.31\pm0.08$ and $\sigma$(mean)/$\sigma$(RMS)$\sim$$1.60\pm0.06$. The continuum variation timescale of Arp 151 is about 80 days (\citealt{Li2013}), much longer than the typical time delay of about 4 days (\citealt{Bentz2009}), indicating that the influence of continuum variation timescale on the line widths is minimized. In other words, the broader line widths of mean spectra in Arp~151 give evidence for a radial dependent responsivity. The treatment with a global constant responsivity $\eta=1$ in dynamical modeling of the BLR is therefore no longer appropriate. It is plausible that some systematics might arise when forcing $\eta=1$ in dynamical modeling and this deserves a detailed investigation in a future work.

\section{Conclusions}\label{sec_conclusion}
Based on linearized response model, we establish a rigorous formulation for RM by accounting for locally dependent echoes of BLRs, inspired from photoionization calculations and the long-standing phenomenon that mean and RMS spectra in RM campaigns are generally different in shape and/or width. To this end, we introduce a radial-dependent responsivity to delineate non-uniform responses of BLRs. This means that some parts of BLRs are more responsive that others, naturally leading to different shapes of mean and RMS spectra. The time series of emission lines is then decomposed into constant and variable components, governed respectively by the emissivity- and responsivity-weighted transfer functions. These two transfer functions are fully determined given a BLR model together with a form of responsivity. Only the variable component can be formulated as a convolution between the transfer function and continuum light curve, similar to the linearized formula used in the maximum entropy method (\citealt{Horne1994}). In our formulation, the mean and RMS spectra can be self-consistently determined, allowing us to explore the virial-based SMBH mass measurements from a theoretical perspective.

Using a simple toy BLR dynamical model, we demonstrate that the observed broader H$\beta$ line widths in mean spectra than in RMS spectra can be explained by a radial increasing responsivity. Through a suite of simulation tests, we find that both line width ratios of mean and RMS spectra and ratios of emissivity- and responsivity-weighted BLR sizes show dependence on specific forms of responsivity. This in turn leads to dependence of the virial products on responsivity, potentially indicating that the virial factors might change with luminosity states of AGNs. Remarkably, due to the spatial extension of BLRs,
the reprocessed variations from larger radii tend to be more strongly blurred and therefore the corresponding time delay information will be effectively filtered out when the driving continuum variations are rapid. As pointed out by \cite{Goad2014}, such a ``geometric dilution effect'' can cause underestimated time delays. We further illustrate that this effect also results in overestimated line widths in RMS spectra and ultimately brings about a systematic bias in virial-based SMBH measurements. The bias can be up to $\sim$0.3 dex in our simulations, but possibly depending on specific BLR dynamical models.

We conclude with two remarks. Firstly, previous dynamical modeling approaches using globally linear or uniform responses of BLRs are inappropriate as this produces completely identical mean and RMS spectra, inconsistent with observations. Systematics might arise in the obtained SMBH masses and BLR sizes.  An updated reanalysis with our framework is highly worthwhile. Secondly, our framework naturally addresses the issue in joint analysis of SA and RM that these two techniques are sensitive to different aspects of BLRs (namely, emissivity and responsivity weights). Therefore, it has a great potential for future joint analysis to better reveal geometry and kinematics of BLRs and also to precisely measure cosmic distances of AGNs.

\section*{Acknowledgements}
We thank the referee for the useful comments that improved the clarity of the manuscript.
We acknowledge financial support from the National Key R\&D Program of China (2023YFA1607904 and 2021YFA1600404), the National Natural Science
Foundation of China (NSFC; 11991050 and 12333003). Y.R.L. acknowledges financial support from the NSFC through grant No. 12273041 and from the Youth Innovation Promotion Association CAS.

\software{\textsc{BRAINS} (\citealt{Li2018, brains}), \textsc{CLOUDY} (\citealt{Gunasekera2023})}

\appendix
\section{Derivations for Reverberation Mapping Equations}\label{sec_app_rm}

By dividing the continuum light curve into constant and variable components and under the condition of small variability, we have an approximation
\begin{equation}
F_c^{\eta}(t) = \left[\bar F_c + \Delta F_c(t)\right]^\eta \approx \bar F_c^{\eta} + \eta\bar F_c^{\eta-1}\Delta F_c(t).
\end{equation}
Substituting this equation into Equation~(\ref{eqn_rm_org}), we have
\begin{eqnarray}
F_l(t, v) & = &  \iint  \bar F_c^{\eta} A(\bm{r})r^{-2\eta} \delta(v-v_{\rm cl})f(\bm{r}, \bm{w}) d\bm{r}d\bm{w}\nonumber\\
&+& \iint  \eta \bar F_c^{\eta-1}\Delta F_c(t-\tau_{\rm cl}) A(\bm{r}) r^{-2\eta} \delta(v-v_{\rm cl})f(\bm{r}, \bm{w}) d\bm{r}d\bm{w},
\end{eqnarray}
where $f(\bm{r},\bm{w})$ is the number distribution of the BLR clouds, $A(\bm{r})$ is a factor defined in Equation~(\ref{eqn_resp1}), $\tau_{\rm cl}$ is the time delay (Equation~\ref{eqn_tauc}), and $v_{\rm cl}$ is the observed velocity (Equation~\ref{eqn_vc}).
It is clear that the two terms in the right-hand side of the above equation correspond to the constant and variable components
of the emission line flux, namely,
\begin{equation}
\bar F_l(v) = \bar F_c\iint \bar F_c^{\eta-1} A(\bm{r})  r^{-2\eta} \delta(v-v_{\rm cl})f(\bm{r}, \bm{w}) d\bm{r}d\bm{w},
\label{eqn_mean}
\end{equation}
and
\begin{equation}
\Delta F_l(t, v) = \iint  \eta \bar F_c^{\eta-1}\Delta F_c(t-\tau_{\rm cl}) A(\bm{r})r^{-2\eta} \delta(v-v_{\rm cl})f(\bm{r}, \bm{w}) d\bm{r}d\bm{w}.
\label{eqn_var}
\end{equation}
By introducing the emissivity-weighted transfer function (e.g., see \citealt{Goad1993})
\begin{eqnarray}
\Psi_{\rm e}(\tau, v) &=& \iint  \bar F_c^{\eta-1}A(\bm{r})r^{-2\eta} \delta(\tau-\tau_{\rm cl})\delta(v-v_{\rm cl})
f(\bm{r}, \bm{w})d\bm{r}d\bm{w},
\end{eqnarray}
and responsivity-weighted transfer function,
\begin{eqnarray}
\Psi_{\rm r}(\tau, v) &=& \iint  \eta \bar F_c^{\eta-1}A(\bm{r})r^{-2\eta} \delta(\tau-\tau_{\rm cl})\delta(v-v_{\rm cl})
f(\bm{r}, \bm{w})d\bm{r}d\bm{w},
\end{eqnarray}
the constant component can be simplified into
\begin{equation}
\bar F(v) = \bar F_c \int \Psi_{\rm e}(\tau, v)d\tau = \Psi_{\rm e}(v) \bar F_c,
\end{equation}
and the variable component of the emission line flux can be  written as a convolution between the responsivity-weighted transfer function
and continuum,
\begin{equation}
\Delta F_l(t, v) = \int \Psi_{\rm r}(\tau, v)\Delta F_c(t-\tau) d\tau.
\label{eqn_rm}
\end{equation}
The above two equations are exactly the same as Equations~(\ref{eqn_rm_mean}) and (\ref{eqn_rm_var}).
The complete expression for the emission-line flux is
\begin{equation}
F_l(t, v) = \Psi_{\rm e}(v)\bar F_c + \int \Psi_{\rm r}(\tau, v)\Delta F_c(t-\tau)d\tau,
\label{eqn_rm_full}
\end{equation}
which encapsulates both the emissivity- and responsivity-weighted transfer functions.

It is worth mentioning that Equation~(\ref{eqn_rm}) is similar to the conventional form of reverberation equation used in previous studies (e.g., \citealt{Peterson1993}). However, it is important to note that the above convolution only involves the variable parts (not the whole) of both emission line and continuum fluxes. Indeed, there is not a simple convolutional expression for the whole emission line and continuum fluxes, except for the case with constant responsivity (see Section~\ref{sec_rms}).
We also stress that the emissivity-weighted transfer function $\Psi_{\rm e}$ is appropriate for calculating the mean spectrum and it does not determine the reverberation and time delay of the BLR.

\section{Derivations for RMS Spectrum}\label{sec_app_rms}
From Equation~(\ref{eqn_rm}), it is straightforward to derive
\begin{eqnarray}
\left[\Delta F_l(t, v)\right]^2 &=& \left[\int \Psi_{\rm r}(\tau, v) \Delta F_c(t-\tau) d\tau \right] \left[\int  \Psi_{\rm r}(\tau', v) \Delta F_c(t-\tau')d\tau'\right]\nonumber\\
&=&\iint \Psi_{\rm r}(\tau, v)  \Psi_{\rm r}(\tau', v) \Delta F_c(t-\tau) \Delta F_c(t-\tau') d\tau d\tau'.
\end{eqnarray}
As a result, the RMS spectrum in Equation~(\ref{eqn_rms_def}) can be rewritten as
\begin{eqnarray}
{\rm RMS} &=& \left[\frac{1}{T}\iiint \Psi_{\rm r}(\tau, v)  \Psi_{\rm r}(\tau', v) \Delta F_c(t-\tau) \Delta F_c(t-\tau') d\tau d\tau' dt\right]^{1/2}.
\end{eqnarray}
Let's define the covariance function of the continuum light curve as
\begin{equation}
S_c(\Delta \tau ) = \frac{1}{T}\int \Delta F_c(t-\tau) \Delta F_c(t-\tau') d t.
\label{eqn_sc}
\end{equation}
where $\Delta \tau=\tau-\tau'$. Colloquially, $S_c(\Delta t)$ represents the auto-correlation of the continuum light curve at a time difference $\Delta \tau$. The RMS spectrum is then recast into
\begin{eqnarray}
{\rm RMS} &=& \left[\iint \Psi_{\rm r}(\tau, v) \Psi_{\rm r}(\tau', v) S_c(\tau-\tau') d\tau d\tau'\right]^{1/2}.
\end{eqnarray}

If the characteristic timescale of the continuum variation is much longer than the typical time delay of the transfer function, the covariance function $S_c(\Delta t)$ can be regarded as a constant and thereby the RMS spectrum is written
\begin{equation}
{\rm RMS} \propto \left[\iint \Psi_{\rm r}(\tau, v) \Psi_{\rm r}(\tau', v) d\tau d\tau'\right]^{1/2} \equiv \Psi_{\rm r}(v).
\end{equation}
On the other hand, if the continuum variation is rapid and the covariance function can be regarded as a delta function $S(\Delta \tau)\propto\delta(\Delta \tau)$, there will be
\begin{equation}
{\rm RMS} \propto \left[\iint \Psi_{\rm r}(\tau, v) \Psi_{\rm r}(\tau', v) \delta(\tau-\tau') d\tau d\tau'\right]^{1/2} \propto \left[\int \Psi_{\rm r}^2(\tau, v) d\tau\right]^{1/2}\equiv \widehat{\Psi}_{\rm r}(v).
\end{equation}

\end{document}